\documentclass[superscriptaddress,twocolumn,showpacs,amsmath,amssymb]{revtex4}

\usepackage{graphicx}

\renewcommand{\vec}[1]{\mbox{\boldmath $#1$}}

\begin{document}

\title{
Two-particle correlations in continuum dipole transitions in 
Borromean nuclei}

\author{K. Hagino}
\affiliation{ 
Department of Physics, Tohoku University, Sendai, 980-8578,  Japan} 

\author{H. Sagawa}
\affiliation{
Center for Mathematical Sciences,  University of Aizu, 
Aizu-Wakamatsu, Fukushima 965-8560,  Japan}

\author{T. Nakamura}
\affiliation{
Department of Physics, Tokyo Institute of Technology, 
2-12-1 O-Okayama, Tokyo 152-8551, Japan}

\author{S. Shimoura}
\affiliation{
Center for Nuclear Study, University of Tokyo (CNS) RIKEN Campus, 2-1 Hirosawa, 
Wako, Saitama 351-0198, Japan}


\begin{abstract}
We discuss the energy and angular distributions of two emitted neutrons 
from the dipole excitation of typical weakly-bound 
Borromean nuclei, $^{11}$Li and $^6$He. 
To this end, we use a three-body model with a density dependent contact 
interaction between the valence neutrons. 
Our calculation indicates that the energy distributions for the valence 
neutrons are considerably different between the two nuclei, 
although they show 
similar strong dineutron correlations in the ground state to each other. 
This different behaviour of the energy distribution 
primarily reflects the interaction between the neutron and the core nucleus, 
rather than the interaction between the valence neutrons. That is, the 
difference
can be attributed to 
the presence of $s$-wave virtual state in 
the neutron-core system in $^{11}$Li, which is absent in $^6$He. 
It is pointed out  that the angular distribution for $^{11}$Li 
in the low energy region shows a clear manifestation of 
the strong dineutron correlation,  
whereas the angular distribution 
for $^{6}$He exhibits a strong anticorrelation effect.   
\end{abstract}


\maketitle

Borromean nuclei are unique three-body bound systems, in which 
any two-body subsystem is not bound \cite{Zhukov93,JRFG04}. 
Typical examples include $^{11}$Li and $^6$He, 
which can be viewed as three-body systems consisting 
of a core nucleus and two valence neutrons. 
The binding energy of these neutron-rich 
nuclei is considerably small (the two-neutron 
separation energy, $S_{\rm 2n}$, is, 378 keV \cite{BAG08} and 975 keV for 
$^{11}$Li and $^6$He, respectively, which can be compared with 
{\it e.g.,} $S_{\rm 2n}=12.2$ MeV for $^{18}$O), and 
a few intriguing features originating from the weakly bound property 
have been found. 
A halo structure, in which the density distribution of valence neutrons 
extends far beyond the core nucleus \cite{T85,HJ87}, and strong 
low-energy electric dipole ($E1$) transition \cite{N06,A99} are 
well-known examples. 

One of the most important current open questions concerning the Borromean 
nuclei is to clarify the characteristic nature of correlations between 
the two valence neutrons, which do not form a bound state in the vacuum. 
A strong dineutron correlation, where the two neutrons take 
a spatially compact configuration, has been theoretically 
predicted for some time \cite{Zhukov93,HJ87,BE91,HS05,HSCS07} 
(see also Refs. \cite{MMS05,PSS07}). 
It has been, however, a difficult task to probe experimentally the dineutron 
correlation. In fact, it is only recently when the strong low-lying dipole 
strength distribution has been observed experimentally in the $^{11}$Li nucleus, 
which strongly suggests the existence of dineutron correlation in this nucleus.  

A more direct information on the dineutron correlation 
may be obtained by 
measuring energy and angular distributions 
of two emitted neutrons \cite{S07,C05,M01}. 
Because neither $^{11}$Li nor $^6$He has a bound excited state, 
the Borromean nuclei have to be broken up to a three-body continuum 
state once they are excited as a result of the interaction with 
another nucleus. 
Notice that the operator which induces the E1 excitation is 
proportional to the center of mass coordinate of the two valence 
neutrons, $\vec{R}=(\vec{r}_1+\vec{r}_2)/2$ \cite{BE91,EB92}, where 
$\vec{r}_1$ and $\vec{r}_2$ are the coordinates for the 
two neutrons. Therefore, the relative motion of the two neutrons, 
$\vec{r}=\vec{r}_1-\vec{r}_2$, is not affected by the E1 excitations at all. 
It is thus interesting to ask 
how the energy and angular distributions 
from the E1 excitation reflect the ground state properties of the 
Borromean nuclei, especially the correlation for the relative motion of 
the neutrons, that is, the dineutron correlation.

The aim of this paper is to address this question theoretically using a 
three-body model for the Borromean nuclei. 
The model which we employ is the same as that in Refs. \cite{EBH97,HS05}, 
that is, the three-body model with a density dependent zero-range 
pairing interaction between the two neutrons. 
The model predicts similar strong dineutron correlations 
for both the $^{11}$Li and $^6$He, although the spin 
structure of di-neutron is somewhat different from each other\cite{HS05}. 
Furthermore, 
the model has successfully reproduced the experimental E1 strength 
distribution for both the nuclei \cite{N06,HS07,EHMS07}. 
Therefore, this three-body model provides a useful means to investigate 
the two-particle 
correlations in the dipole transitions in 
the $^{11}$Li and $^6$He nuclei. 
Moreover, by studying the two neutron correlations, 
one may also be able to shed some light on the 
Efimov effect \cite{GFJ06}, that is a general feature of a three-body 
system in which at least two of the three two-body subsystems have an 
infinite $s$-wave scattering length \cite{E70}. 

The energy and angular 
distributions based on 
the three-body model with a density dependent 
contact interaction have been computed 
by Esbensen and Bertsch in Ref. 
\cite{EB92}. As all the basic formulas can be found there, 
we do not repeat them here. 
For the neutron-neutron and the neutron-core interactions, 
we use exactly the same parameters as those in 
Ref. \cite{HS05}, except for the radius parameter for the 
density dependent term in the pairing interaction for the $^{11}$Li nucleus. 
We have slightly adjusted it so that the new empirical value of 
$S_{\rm 2n}$=378keV \cite{BAG08}
is reproduced. This yields the $s$-wave probability of 20.6\% in 
the ground state of $^{11}$Li. 
In order to calculate the continuum response, we 
treat approximately 
the recoil kinetic energy of the core nucleus for the three-body final 
state of the dipole response, although it is treated exactly for the 
initial (ground) state. 
That is, we ignore 
the off-diagonal
component (the last term in Eq. (3.3) of
Ref. \cite{EBH97}), whereas the diagonal term 
is included through the reduced mass \cite{HS07,EHMS07}. 
We have checked
the accuracy of this approximation with the discretized dipole
strength functions and have confirmed that the
approximation works well both for $^{11}$Li and $^6$He \cite{HS07}. 


\begin{figure}[htb]
\hspace{-1cm}
\includegraphics[clip,scale=0.85]{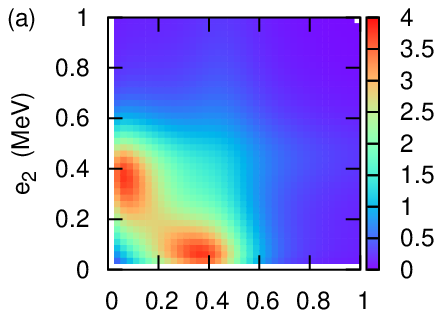}
\hspace{-1.5cm}
\includegraphics[clip,scale=0.85]{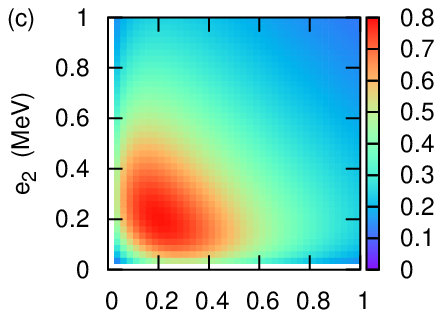}\\
\vspace{-1.15cm}
\hspace{-1cm}
\includegraphics[clip,scale=0.85]{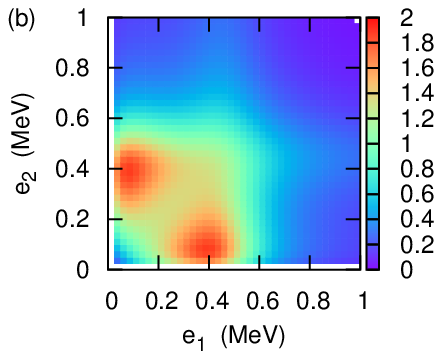}
\hspace{-1.5cm}
\includegraphics[clip,scale=0.85]{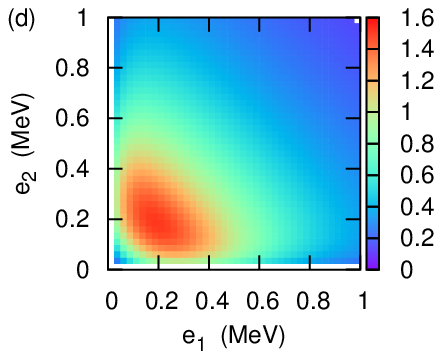}
\caption{(Color online)
The dipole strength distributions, $d^2B(E1)/de_1de_2$, of 
$^{11}$Li as a function of
the energies of the two emitted neutrons relative to the core nucleus. 
They are plotted in units of $e^2$fm$^2$/MeV$^2$. 
Fig. 1(a) shows the correlated response, which 
fully takes into account the ground state 
 and final state interactions between the 
two neutrons, while Fig. 1(b) shows the unperturbed response, obtained 
by neglecting the neutron-neutron interaction in the final states. 
Figs. 1(c) and 1(d) are 
obtained by assuming the plane-wave for the final 
states ({\it i.e.,} neglecting the neutron-core interaction)  
without and with the final state interaction between the 
two neutrons, respectively. 
}
\end{figure}

Figures 1 and 2 show the dipole strength distribution, $d^2B(E1)/de_1de_2$, 
as a function of the energies of the two emitted neutrons for the 
$^{11}$Li and 
$^6$He nuclei, respectively. Here, $e_1$ ($e_2$) is the relative energy between 
the first (second) neutron and the core nucleus. Notice that these energy distributions 
are symmetric with respect to the interchange of $e_1$ and $e_2$. 
Fig. 1(a) shows the correlated response, which 
fully takes into account the final state interaction between the 
two neutrons, while Fig. 1(b) shows the unperturbed response, obtained 
by neglecting the neutron-neutron interaction in the final states. 
Figs. 1(c) and 1(d) are 
obtained by assuming the plane-wave for the final 
states ({\it i.e.,} neglecting the neutron-core interaction)  
without and with the final state interaction between the 
two neutrons, respectively. 
Figs. 2(a)-2(d) are the same, but for $^{6}$He.  
All the calculations are performed by using the Green's function 
method \cite{EB92}. 

One immediately notices that the strength distribution is considerably 
different between $^{11}$Li and $^6$He. For $^{11}$Li, a large concentration 
of the strength appears at about $e_1$=0.375 MeV and $e_2$=0.075 MeV 
(and at $e_1$=0.075 MeV and $e_2$=0.375 MeV), with 
a small ridge at an energy of about 0.5 MeV. 
On the other hand, for $^6$He, only a large ridge at about 0.7 MeV appears, 
and the strength is largely concentrated around $e_1=e_2=0.7$ MeV. 
These features remain the same even if the final state interaction between the 
two emitted neutrons is switched off, as shown in Figs. 1(b) and 2(b), 
although the degree of the concentration of the strength is much more 
emphasized by the final state interaction. 
In contrast, if the interaction between the neutron and the core nucleus 
is neglected, the strength distribution is altered drastically, 
and in fact the distribution is now 
similar between the two nuclei (see Figs. 1(c), 1(d), 2(c), and 2(d)). 
Therefore, the different behaviours in the strength distribution should 
reflect primarily the property of the neutron-core interaction. 
In fact, the comparisons between Figs. 1(a) and 1(b), 1(c) and 1(d), 
2(a) and 2(b), and 2(c) and 2(d) indicate that the final state 
 neutron-neutron interaction 
does not play a major role in the shape of energy distribution. 


\begin{figure}[htb]
\hspace{-1cm}
\includegraphics[clip,scale=0.85]{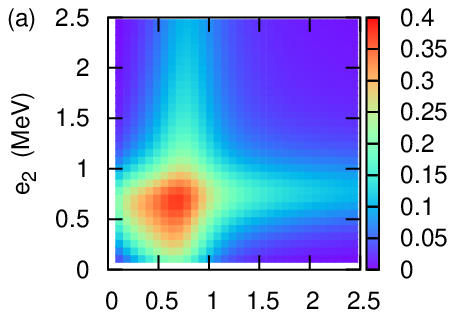}
\hspace{-1.5cm}
\includegraphics[clip,scale=0.85]{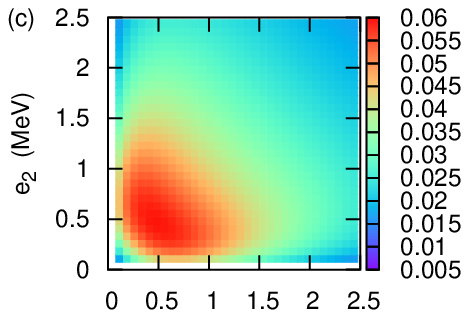}\\
\vspace{-1.15cm}
\hspace{-1cm}
\includegraphics[clip,scale=0.85]{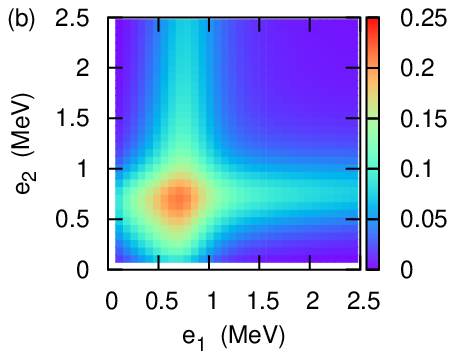}
\hspace{-1.5cm}
\includegraphics[clip,scale=0.85]{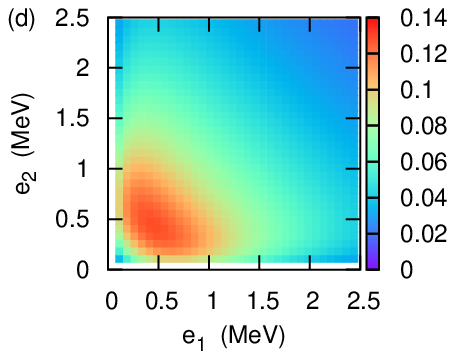}
\caption{(Color online)
Same as Fig.1, but for $^6$He. 
}
\end{figure}

The ridges in the strength function have already been discussed in the 
previous calculation for $^{11}$Li by Esbensen and Bertsch \cite{EB92}. 
They reflect the single-particle resonances, that is, the p$_{1/2}$ resonance 
around 0.54 MeV for $^{10}$Li and the p$_{3/2}$ resonance at 0.91 MeV 
for $^5$He \cite{EBH97}. These ridges correspond to the physical 
process in which one of the neutrons is excited by the dipole field while 
the other remains near the resonance state as a spectator in the 
neutron-core system \cite{BE91,EB92}. 
For $^{11}$Li, in addition to the ridge, the dipole strength is concentrated 
in the region in which one of the neutrons has an energy close to zero. 
This reflects the $s$-wave virtual state in $^{11}$Li close to 
zero energy\cite{EBH97,GFJ06,Y94,TZ94}.  
This virtual state is characterized by a large negative scattering length of 
$a=-30^{+12}_{-31}$ fm \cite{S07} for the $n+^9$Li system. 
In contrast, the $s$-wave scattering length is $a=4.97\pm 0.12$ fm 
\cite{A73} for the $n+^4$He system, and the virtual state does not 
exist in $^5$He. Notice that the virtual state was not taken into account 
in the previous calculation for the dipole response of $^{11}$Li, and 
the concentration of the strength in the region of $e\sim$ 0 
was not found in Ref. \cite{EB92}. 

\begin{figure}[htb]
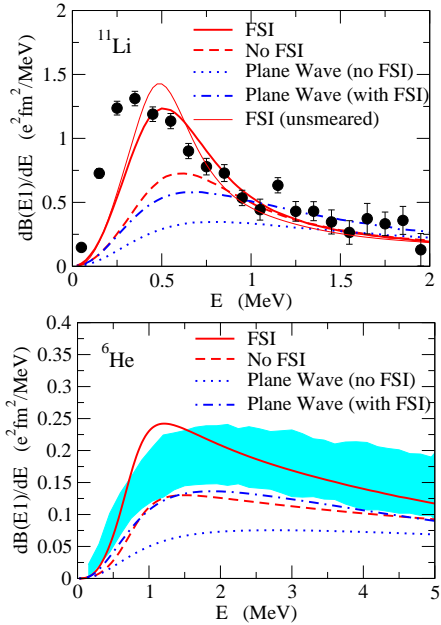

\includegraphics[clip,scale=0.4]{fig3-a}\\
\includegraphics[clip,scale=0.4]{fig3-b}
\caption{(Color online)
The upper panel: 
The dipole strength distribution for $^{11}$Li 
as a function of $E=e_1+e_2$. 
The thick solid, dashed, dotted, and dot-dashed lines 
are obtained under the same assumptions for 
 Figs. 1(a), 1(b), 1(c), and 1(d), respectively. 
These curves are smeared with the 
experimental energy resolution. 
The thin solid line is the same as the thick solid line, but 
without the smearing. 
The experimental data are taken from \cite{N06}. 
The lower panel: the same as the upper panel, but for 
$^{6}$He. The shaded area shows the experimental data, taken from 
Ref. \cite{A99}. 
}
\end{figure}

The dipole strength distribution for $^{11}$Li 
at a fixed total energy of around $E=e_1+e_2$=0.45 MeV 
has two distinguished peaks, one at $e_1/E\sim 0$ 
and the other at $e_1/E\sim 1$. 
It has been argued that the two peaked structure in the energy 
distribution is a characteristic feature of the Efimov effect \cite{GFJ06}. 
It is a peculiar three-body effect for a system in which two or three 
two-body subsystems have a large scattering length. 
In $^{11}$Li, this condition is approximately fulfilled, 
as the scattering length is large and negative both for the $nn$ and 
the $n$-core systems. 
In our calculation, the $s$-wave scattering length is $-$5.6 fm \cite{EBH97} 
for the $n$-$^9$Li and +2.43 fm for the $n$-$^4$He systems, where as it is 
$-$15 fm for the $n-n$ system\cite{EBH97}. 

\begin{figure}[htb]
\hspace{-1cm}
\includegraphics[clip,scale=0.85]{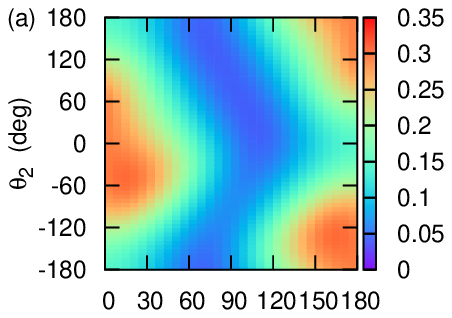}
\hspace{-1.5cm}
\includegraphics[clip,scale=0.85]{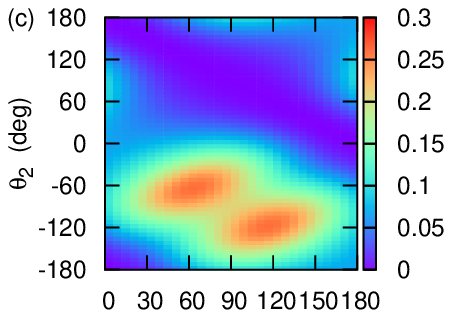}\\
\vspace{-1.15cm}
\hspace{-1cm}
\includegraphics[clip,scale=0.85]{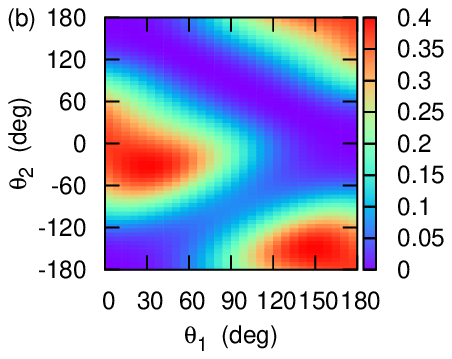}
\hspace{-1.5cm}
\includegraphics[clip,scale=0.85]{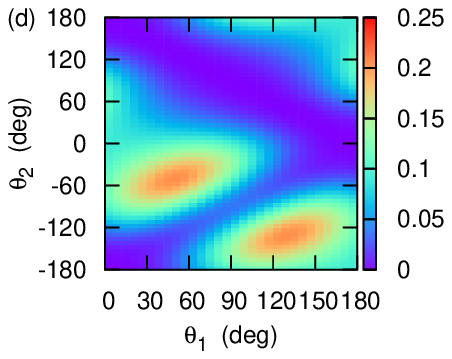}
\caption{(Color online)
Angular distributions of the two valence neutrons 
in $^{11}$Li and $^6$He 
emitted in the rest frame of the corresponding nuclei. 
These are calculated 
for the configuration in which the two neutrons are emitted 
in the same reaction plane ({\it i.e.,} $\phi_1=\phi_2$) 
by the longitudinal 
component of the E1 operator. 
Fig. 4(a) shows the angular distribution for $^{11}$Li 
at $e_1=0.375$ MeV and $e_2=0.075$ MeV, while 
Figs. 4(b) and 4(c) are for 
$e_1=e_2=0.225$ MeV 
and $e_1=e_2=0.5$ MeV, respectively. 
Fig. 4(d) shows the angular distribution for $^{6}$He 
at $e_1=e_2=0.7$ MeV. 
}
\end{figure}

The dipole strength distributions, 
\begin{equation}
\frac{dB(E1)}{dE}=\int de_1de_2 \,
\frac{d^2B(E1)}{de_1de_2}\,\delta(E-e_1-e_2)\,,
\end{equation}
are plotted in Fig. 3. The solid, dashed, dotted, and dot-dashed 
lines are obtained under the same assumptions for 
Figs. 1(a)/2(a), 1(b)/2(b), 1(c)/2(c), 
and 1(d)/2(d), respectively. 
These curves are smeared with the 
experimental energy resolution \cite{N06}. 
Also shown by the thin solid line in Fig. 3(a) 
is the same as the thick solid line, but without the smearing . 
The experimental B(E1) distribution for $^{11}$Li 
is obtained from the experimental breakup cross sections in Ref.\cite{N06}, 
using $S_{2n}$=378 keV. 
Because of the large concentration of the strength in the 
low energy region in the energy distribution shown in Fig. 1, 
the strength distribution 
for $^{11}$Li has a significantly sharper peak as compared to that for $^6$He. 
Notice again that the strength distributions are qualitatively 
similar between the two 
nuclei if the neutron-core potential is neglected (see the dotted and the 
dot-dashed lines). 

In order to discuss how the dineutron correlation in the ground state 
affects the dipole response, let us next consider the angular 
distributions of the two emitted neutrons. 
Figure 4 shows the angular distributions 
corresponding to 
the longitudinal component of the dipole excitation of 
the $^{11}$Li and $^6$He nuclei 
(induced by the $\mu=0$ component of the E1 operator, 
$\hat{T}_{\lambda=1,\mu=0}$), that is, $\sum_{h_1,h_2}|
f^{10}_{h_1h_2}(\hat{\vec{k}}_1,\hat{\vec{k}}_2)|^2$, where $h$ and 
$\hat{\vec{k}}=(\theta,\phi)$ 
are the helicity and the direction of the momentum vector 
for an emitted neutron, respectively. 
Here, 
$f^{1\mu}_{h_1h_2}(\hat{\vec{k}}_1,\hat{\vec{k}}_2)$ 
is the amplitude of dipole excitation calculated with 
two-particle wave functions in the continuum state with 
definite momenta $\vec{k}_1$ and $\vec{k}_2$ and 
definite spins, where the momenta are defined in the rest frame 
of $^{11}$Li and $^6$He (see Eq. (5.5) in Ref. \cite{EB92}). 

The angular distribution for $^{11}$Li 
for the energy at which the dipole strength 
is concentrated, that is, $e_1=0.375$ MeV and $e_2=0.075$ MeV (see Fig. 1), 
is shown in Fig. 4 (a). 
The distribution is 
calculated for the configurations in which the two emitted neutrons 
are in the same reaction plane, that is, $\phi_1=\phi_2$. 
Here the second neutron mainly occupies the $s$-wave virtual state, 
and the angular distribution for this neutron is widely spread. 
The energy and angular distributions are mainly determined by the virtual 
state for this 
energy configuration, and the effect of final state interaction 
does not seem to play a major role, as one can infer from Fig. 1. 
We have confirmed this by switching off the final state interaction 
in our calculation. We have in fact obtained a 
qualitatively similar angular 
distribution as in Fig. 4 (a). 

The angular distribution for the same total energy $e_1+e_2=0.45$ MeV as 
in Fig. 4(a), but for $e_1=e_2=0.225$ MeV, is plotted in Fig. 4 (b). 
For this configuration, 
both the neutrons are 
emitted along the $z$ axis ({\it i.e.,} $\theta_1=\theta_2=0$) 
with a large probability, although the distribution is rather flat around 
$\theta_1=\theta_2=0$, with the maximum at $\theta_1=-\theta_2=30^{\circ}$. 
Therefore, the opening angle between the two neutrons is relatively small, and 
one would naively consider that 
the shape of this distribution strongly reflects the dineutron 
correlation, with some 
perturbation by the anticorrelation effect for the dipole excitation. 
As noted in Ref. \cite{EB92}, the anticorrelation, with which the two 
neutrons are emitted on opposite sides of the $z$ axis, is associated with 
 the projection of the wave function with a coherent mixture of 
$s$ and $d$ wave states onto the momentum states.

The anticorrelation
is more pronounced at higher energies. Fig. 4 (c) shows the 
angular distribution at 
$e_1=e_2=0.5$ MeV, which corresponds to the region of the ridge in the energy 
distribution shown in Fig. 1. 
For this energy configuration, the probability for emission of two neutrons 
on the same sides of the $z$ axis 
(in the region of $\theta_2 > $0) is largely suppressed, and the maximum of 
the distribution appears around 
$\theta_1=60^\circ$ and $\theta_2=-66^\circ$. Notice that the shape of the 
distribution is similar to 
the results of the previous calculation shown in Figs. 9 and 10 in 
Ref. \cite{EB92}. 
The shape is determined by a destructive interference between 
the [d$\otimes$p] and 
[p$\otimes$s] configurations \cite{EB92}, excited from the 
[p$^2$] 
and [s$^2$] 
configurations in the ground state wave function. 
Therefore, the angular distribution around this energy 
is strongly affected by the 
$p$-wave single particle resonance.
In fact, for the $^6$He nucleus, where the energy distribution is characterized  by the p$_{3/2}$ resonance, we obtain a very similar angular distribution 
at $e_1=e_2=0.7$ MeV (see Fig. 4(d)). 
For $^6$He, it would therefore not be straightforward to probe the dineutron 
correlation in the ground state solely by the 
angular distribution, in which the dineutron correlation is largely masked 
by the effect of the $p$ wave single particle resonance. 

In summary, we have carried out the three-body model calculations for 
$^{11}$Li and $^6$He nuclei in order to investigate the energy and angular 
distributions of the two emitted neutron from E1 excitations. 
We have shown that these distributions are strongly 
affected by the properties of 
the neutron-core potential, rather than the ground state properties. 
For the $^{11}$Li nucleus, the presence of $s$-wave virtual state 
helps to reveal a clear manifestation of the strong 
dineutron correlation through the energy and 
the angular distributions. For the $^6$He nucleus, on the other hand, 
it is not straightforward to  probe it,  due to  the anticorrelation effect 
  in the angular distribution. 
The correlation measurements for $^{11}$Li have been recently done 
at RIKEN. 
The present calculations shown in this paper
are qualitatively in good agreement with the 
preliminary data. 
We will report the analyses of these 
data in a separate paper. 


We thank P. Schuck for useful discussions. 
This work was supported by the Japanese
Ministry of Education, Culture, Sports, Science and Technology
by Grant-in-Aid for Scientific Research under
the program numbers (C) 20540277 and 19740115.

\end{document}